\documentclass[epj]{svjour}
\usepackage{graphicx}
\usepackage{verbatim}
\usepackage{amsmath}
\usepackage{amssymb}
\usepackage{epstopdf}
\usepackage{color}

\newcommand{\sqrts}{\sqrt{s}}
\newcommand{\snn}{\sqrt{s_{\rm NN}}}
\newcommand{\pt}{p_{\rm T}}
\newcommand{\raa}{R_{\rm AA}}

\newcommand{\jpsi}{$J/\psi$}

\begin{document}

\title{Heavy-flavor production and medium properties in high-energy nuclear collisions -- What next?}

\author{
G. Aarts\inst{1} \and
J. Aichelin (co-organiser)\inst{2}  \and
C. Allton\inst{1}  \and
R. Arnaldi (co-convener)\inst{3} \and
S. A. Bass\inst{4} \and
C. Bedda\inst{5} \and 
N. Brambilla\inst{6, 7} \and
E. Bratkovskaya\inst{8, 9} \and
P. Braun-Munzinger\inst{8, 10} \and 
G. E. Bruno\inst{11} \and
T. Dahms (co-convener)\inst{6} \and
S. K. Das\inst{12} \and
H. Dembinski\inst{13} \and
M. Djordjevic\inst{14} \and
E. Ferreiro\inst{15} \and
A. Frawley\inst{16} \and
P.-B. Gossiaux\inst{2} \and
R. Granier de Cassagnac (co-organiser)\inst{17} \and
A. Grelli\inst{5} \and
M. He\inst{18} \and
W. A. Horowitz (co-convener)\inst{19} \and
G. M. Innocenti\inst{20} \and
M. Jo\inst{17} \and
O. Kaczmarek\inst{21, 22} \and
P. G. Kuijer\inst{23} \and
M. Laine\inst{24} \and
M. P. Lombardo (co-organiser)\inst{25} \and
A. Mischke (co-organiser)\inst{5, 23} \and
M. G. Munhoz\inst{26} \and
M. Nahrgang\inst{2} \and
M. Nguyen (co-convener)\inst{17} \and 
A. C. Oliveira da Silva\inst{26, 5} \and
P. Petreczky (co-convener)\inst{27} \and
A. Rothkopf\inst{28} \and
M. Schmelling\inst{13} \and
E. Scomparin\inst{3} \and
T. Song\inst{9} \and
J. Stachel\inst{10} \and
A. A. P. Suaide\inst{26} \and
L. Tolos\inst{29, 30} \and
B. Trzeciak\inst{5} \and
A. Uras\inst{31} \and
L. van Doremalen\inst{5} \and
L. Vermunt\inst{5} \and
S. Vigolo\inst{5} \and
N. Xu (co-organiser)\inst{32} \and 
Z. Ye\inst{33} \and
H. J. C. Zanoli\inst{26, 5} \and
P. Zhuang\inst{34}
}
%
\institute{
Swansea University, Swansea, United Kingdom \and
SUBATECH, \`Ecole des Mines de Nantes, Universit\'e de Nantes, CNRS-IN2P3, Nantes, France \and
INFN Sezione, Torino, Italy \and
Duke University, Durham, USA \and
Institute for Subatomic Physics, Utrecht University, Utrecht, the Netherlands \and
Physik-Department and Excellence Cluster Universe, Technische Universit\"at M\"unchen, Garching, Germany \and
Institute for Advanced Study, Technische Universit\"at M\"unchen, Munich, Germany \and
Research Division and ExtreMe Matter Institute EMMI, GSI Helmholtzzentrum f\"ur Schwerionenforschung, Darmstadt, Germany \and
Institute for Theoretical Physics, Frankfurt University, Frankfurt am Main, Germany \and
Physikalisches Institut, Ruprecht-Karls-Universit\"at Heidelberg, Heidelberg, Germany \and
Dipartimento di Fisica and INFN, Bari, Italy and European Organization for Nuclear Research, Geneva, Switzerland \and
University of Catania, Catania, Italy \and
Max-Planck-Institut f\"ur Kernphysik, Heidelberg, Germany \and
Institute of Physics, University of Belgrade, Serbia \and
Universidad de Santiago de Compostela, Santiago de Compostela, Spain \and
Florida State University, Tallahassee, USA \and
Laboratoire Leprince-Ringuet, \`Ecole polytechnique, France \and
Department of Applied Physics, Nanjing University of Science and Technology, Nanjing, China \and
Department of Physics, University of Cape Town, Rondebosch, South Africa \and
Massachusetts Institute of Technology, USA \and
Key Laboratory of Quark \& Lepton Physics (MOE) and Institute of Particle Physics, Central China Normal University, Wuhan, China \and
University of Bielefeld, Bielefeld, Germany \and
Nikhef, National Institute for Subatomic Physics, Amsterdam, the Netherlands \and
AEC, Institute for Theoretical Physics, University of Bern, Switzerland \and
INFN - Laboratori Nazionali di Frascati, Italy \and
Universidade de S\~{a}o Paulo (USP), S\~{a}o Paulo, Brazil \and
Brookhaven National Laboratory, Upton, USA \and
Institute for Theoretical Physics, Ruprecht-Karls-Universit\"at Heidelberg, Heidelberg, Germany \and
Frankfurt Institute for Advanced Studies (FIAS), Frankfurt, Germany \and
Institut de Ci\`encies de l'Espai, (IEEC-CSIC), Bellaterra, Spain \and
Institute of Nuclear Physics, Domaine scientifique de la Doua, Villeurbanne Cedex, France \and
Lawrence Berkeley National Laboratory, Berkeley, USA \and
University of Illinois, Chicago, USA \and
Tsinghua University, China 
}

\date{\today}

\abstract{
Open and hidden heavy-flavor physics in high-energy nuclear collisions are entering a new and exciting stage towards reaching a clearer understanding of the new experimental results with the possibility to link them directly to the advancement in lattice Quantum Chromo-dynamics (QCD). 
Recent results from experiments and theoretical developments regarding open and hidden heavy-flavor dynamics have been debated at the Lorentz Workshop {\em Tomography of the quark-gluon plasma with heavy quarks}, which was held in October 2016 in Leiden, the Netherlands. 
In this contribution, we summarize identified common understandings and developed strategies for the upcoming five years, which aim at achieving a profound knowledge of the dynamical properties of the quark-gluon plasma.
}



\maketitle


\section{Introduction}
Over the last decade, different experimental observables have been used for the characterization of the quark-gluon plasma (QGP)~\cite{Brambilla:2014jmp}. Heavy quarks play a crucial role as a probe thanks to their large mass with respect to the temperature of the plasma consisting of gluons and light quarks. Therefore, heavy quarks are ideal probes for the study of the QGP properties~\cite{hfGeneral,Brambilla:2010cs} because they are produced in the very early stage of the collision testing the entire space-time evolution of the system. 
The availability of a heavier (bottom) and lighter (charm) flavor offers the unique possibility to probe different stages of this space-time evolution. For bottom the thermalization time is likely to be larger than the lifetime of the plasma, so that such a non-fully thermalized probe can carry information starting from the earliest moments after its initial creation. For charm on the other hand there is an increasing number of experimental indications for a high degree of equilibration. This in turn means that most information on the evolution history is lost and the late stages around freeze-out dominate the observed behavior.
In addition, from the theoretical point of view, the large mass of heavy quarks makes the evaluation of the so-called quarkonium correlators and transport coefficients feasible directly from first principle QCD calculations.

The experimental results from the Relativistic Heavy Ion Collider (RHIC) and the Large Hadron Collider (LHC) have surprisingly shown a large suppression of the transverse momentum dependent nuclear modification factor $\raa$ of heavy-quark hadrons, which is defined as the ratio of the yields in AA and pp collisions, scaled by the averaged number of binary nucleon-nucleon collisions. In addition, a large heavy-flavor elliptic flow $v_2$ has been observed in heavy-ion collisions. 
This puts additional pressure on theoretical models to reproduce both properties at the same time. Similar to the developments in the area of photon yields and flow, this fruitful challenge already helps theorists to gain a better understanding of the relevant physical processes required in their models of quarkonium production in heavy-ion collisions.

While the measurements of the dynamics of heavy quarks in the medium became feasible in the last decade, the physics of quarkonium production and dissociation is historically one of the main probes of the existence of the QGP and has been studied for nearly thirty years. The new experiments at the LHC and their relation to the results from RHIC and SPS allow clarifying the expected quarkonium melting along with the recombination and regeneration dynamics in the plasma. Moreover, new insights were obtained from the recent developments in lattice QCD from the evaluation of the spectral functions and the possibility at the LHC to reconstruct experimentally the presence of single excited states in the QGP for bottomonium states. This is opening up the possibility to have stringent constraints from both the theoretical and experimental sides for the understanding of the quarkonium production in the plasma.

The Lorentz workshop {\em Tomography of the quark-gluon plasma with heavy quarks}~\cite{LorCenWWW}, which was held on 10-14 October 2016 in Leiden, the Netherlands, provided a platform 
to discuss recent results from experiments and theoretical developments regarding open and hidden heavy-flavor dynamics in high-energy nucleus-nucleus collisions.
Three Discussion Groups were set up to debate in detail among the experts implications and open issues concerning the theoretical and experimental results. They are centered around the following broader questions:
\begin{itemize}
\item Which of the proposed energy-loss mechanisms are compatible with the present lattice results? 
\item What are the next steps for the comparison of the different models for the heavy-quark energy loss in the QGP? 
\item What are the current crucial experimental issues and limitations? Can we identify key observables?
\end{itemize}
The paper gives a summary of the main conclusions and recommendations of the Discussions Groups.

\section{Challenges in QCD theory related to heavy-flavor probes (Discussion Group 1)} \label{sec:dg1}
%
Theoretical efforts (cf.\ e.g.\ refs.~\cite{Brambilla:2014jmp,hfGeneral,Brambilla:2010cs,Beraudo:2014boa,Mocsy:2013syh,Ding:2015ona} 
for reviews) discussed in the Discussion Group may be broadly grouped in two classes:  
\begin{itemize}
  \item First principles calculations of the equilibrium (static and real-time) properties of the QGP from QCD.
  \item Connection of such properties to phenomenological models and experimental results, which requires 
  relating equilibrium processes to non-equilibrium ones (through linear response theory or beyond).
\end{itemize}

The challenges in the first area  concern e.g.\
efficient ways to evaluate observables using different approaches, such as resummed 
perturbation theory, effective field theories  or lattice QCD calculations. The latter area 
on the other hand requires efforts to  link phenomenological ideas, such as
a momentum dependent diffusion coefficient for heavy quarks or an effective transport coefficient
$\hat{q}$ to theoretically well-defined observables. This second step requires close 
interactions among theorists, model builders and experimentalists. 
The joint sessions with other discussion groups focused mostly on these aspects. 

\subsection{Calculations of the equilibrium properties of the quark-gluon plasma from QCD}
Among the observables, which characterize the physics of the QGP, we considered the 
following areas:  

\subsubsection{Bulk thermodynamics}
Lattice QCD computations have converged on an equation of state~\cite{Ratti:2016lrh} 
in the range of 150-300 MeV for a realistic QCD medium
with light $u$, $d$ and $s$  quarks. They can thus now reliably be used as input to
hydrodynamic descriptions of the bulk evolution in heavy-ion collisions~\cite{McDonald:2016vlt}, 
which may still be based on older parameterizations~\cite{Huovinen:2009yb}. 
Recently, studies incorporating dynamical charm quarks
have also seen progress and showed that the effects of charm already set in
at temperatures as low as 400~MeV~\cite{Borsanyi:2016ksw}, provided that the 
plasma lives long enough to reach chemical equilibrium, which may be 
the case in future generations of heavy-ion collision experiments~\cite{Dainese:2016gch}.

The lattice can also study thermodynamic fluctuations and correlations with an
emphasis on charm-quark physics \cite{Kaczmarek:2016xix,Bazavov:2014yba}. 
An interesting outcome is that charm fluctuations and correlations, 
which are sensitive to the open-charm meson and charmed baryon sector,
are well described by hadron resonance gas below $T_{\rm c}$.
%
Above the transition temperature the hadron resonance gas description is not
adequate. At the same time, the lattice calculations of charm
fluctuations and correlations indicate possible existence of charm meson
and baryon like resonances in an extended temperature region $(T\lesssim$ 200~MeV)
\cite{Mukherjee:2015mxc}. These hadronic like excitations are different from the
vacuum charm hadrons and their presence above $T_{\rm c}$ should be taken into account
in the phenomenological models aiming at the description of heavy-flavor production.
For temperatures $T > 300$~MeV the description of fluctuations and correlations
in terms of weak coupling calculations appears to be sufficiently 
accurate~\cite{Bellwied:2015lba,Ding:2015fca,Bazavov:2013uja}. Therefore, models based 
on solely quark quasi-particles are appropriate in this temperature range.

To  make further progress more accurate lattice results on fluctuations and correlations
of charm are needed. Furthermore, quasi-particle models, which include both quark and 
hadron like degrees of freedom above $T_{\rm c}$ that fit the lattice results need to be developed. 
This will ensure to get adequate understanding of the relevant degrees of freedom in
the charm sector.

\subsubsection{Quarkonium and baryon spectroscopy}
Progress on QGP spectroscopy takes place on several fronts. 
The main distinction concerns whether the quarks are light or heavy, 
with charm being intermediate. Spectral information is contained in 
temporal correlation functions; since the temporal extent of the
Euclidean time direction is inversely proportional to the temperature,
higher temperatures are more difficult to analyze.

A complementary way to study in-medium properties of various
excitations comes from the analysis of spatial correlation functions. 
They are less sensitive to e.g.\ transport but provide information
on dissolution of bound states and chiral symmetry restoration (for light
quarks). The advantage is that they are not constrained by the finite
temporal extent, but the disadvantage is the less direct relation with
phenomenological questions~\cite{Bazavov:2014cta}. \newline 

\noindent{\bf Spectral reconstructions using Bayesian inference}\\
The reconstruction of spectral functions from Euclidean correlation functions
represents an ill-posed inverse problem~\cite{Jarrell:1996rrw}. As the number of available simulation points 
is usually much less than the number of frequency bins in which the spectrum is discretized,
naive $\chi^2$ fits will lead to degenerate sets of spectra that all reproduce the
correlators within statistical errors. In order to give meaning to the inversion additional
information needs to be provided and the Bayes theorem may be used to
systematically include such ``priors''. This is achieved by introducing a regulator functional,
which competes with the usual $\chi^2$ fitting functional in order to select the ``most probable''
spectrum.

Bayesian approaches to spectral function computation, such as the Maximum
Entropy Method (MEM)~\cite{Asakawa:2000tr} or the more recent Bayesian
Reconstruction (BR)~\cite{Burnier:2013nla}, differ both due to the regulator
term used as well as in their numerical implementation. Importantly, two
different implementations may give different results as long as the ``Bayesian
continuum limit'' (infinite number of data points, vanishing statistical
errors) has not been taken. In this limit, the problem is well-posed~\cite{Cuniberti:2001hm} 
and all methods should agree, but in practice the limit is far from being reached. 

Over the past two years the systematic artifacts of the two methods have been
much better understood, in particular in cases where only a small number of
data points is available, e.g.\ on the lattice at high temperatures. The
standard implementation of the MEM introduces a limitation to the space of
functions among which the spectrum may be chosen~\cite{Bryan}. This may lead
to an overly smooth reconstruction~\cite{Rothkopf:2011ef}. One way of testing
this limitation is by changing the size of the search space, e.g.\ via the
number of data points included. This is the approach followed by the FASTSUM 
collaboration~\cite{Aarts:2016nwb,Aarts:2014cda}. The BR method on the other 
hand uses a different regulator and does not restrict the search space {\it a priori}. In
turn, it may suffer from the appearance of numerical ``ringing'' that can mimic
the presence of spectral peaks. Here, the comparison of reconstructions using
test cases where peaked structures are absent, e.g.\ non-interacting spectra
have been used as test cases~\cite{Kim:2014iga}.

Differences in the outcomes of the two methods can only be resolved as we
proceed towards the Bayesian continuum limit. Several groups are actively
working on increasing the statistics of the underlying data sets and/or
generating lattices with more finely spaced Euclidean time axes.

It may be noted that non-Bayesian approaches, such as Tichonov-Morozov
\cite{Dudal:2013yva} and Backus-Gilbert~\cite{Brandt:2015sxa}, as well as 
constraints following from the analyticity properties of the   
underlying correlation functions~\cite{Ferrari:2016snh}, 
have recently gained attention. They do not contain an explicit use of
prior information, even if certain ``regulators'' are needed in practice. \newline 

\noindent{\bf Quarkonium spectral functions from NRQCD with full relativistic light quarks}\\
Heavy quarks may be treated within a sequence of effective field theories.
There are several lattice groups using NRQCD (non-relativistic QCD) to
treat bottom quarks propagating through a quark-gluon plasma with
$N_{\rm f} = 2+1$ dynamical (i.e.\ fully relativistic) light flavors~\cite{Aarts:2016nwb,Aarts:2014cda,Kim:2014iga}. 
In the effective thermal field theory setup, NRQCD is the first theory obtained when 
integrating out ultraviolet degrees of freedom.  Since NRQCD relies on the scale separation 
between the temperature and the heavy-quark mass and temperatures up to 
$2T_{\rm c} \simeq 400$~MeV are studied, its application is fully justified.
While one group uses lattices with a very fine temporal spacing 
($a_\tau\simeq 0.035$ fm, $a_s = 3.5a_\tau$) their light medium degrees of freedom 
such as pions are heavier than in nature~\cite{Aarts:2016nwb,Aarts:2014cda}. 
The other group utilizes lattices designed to provide a realistic medium with almost 
physical pion mass but on which the spatial lattice spacing may become very fine 
($0.07 \lesssim a_{\rm s} = a_\tau \lesssim 0.12$~fm), which limits the validity of the 
NRQCD approximation~\cite{Kim:2014iga}.

At low temperatures and for surviving bound states in the QGP, in
particular the $\Upsilon$(1S), groups are in approximate agreement, although
uncertainties on the width are still present. For other channels, notably P-wave 
states within the QGP, this is not yet the case. \newline 

\noindent{\bf Heavy quarkonium spectral functions from pNRQCD}\\
The effective field theory (EFT) for quarkonium at the scale of the 
relative momentum transfer $m v$, namely pNRQCD (potential 
non-relativistic QCD, a lower energy version of NRQCD) has also 
been generalized to a thermal environment~\cite{Brambilla:2008cx}. 
For tightly bound quarkonia like $\Upsilon$(1S) it can be used
to calculate in-medium meson properties and  
the quarkonium thermal width~\cite{Brambilla:2010vq}, induced
by an imaginary part in the in-medium 
potential~\cite{Brambilla:2008cx,Laine:2006ns,Beraudo:2007ky}. 
Such an imaginary part can be related to gluo-dissociation and inelastic 
parton scattering in the medium~\cite{Brambilla:2010vq,Brambilla:2011sg,Brambilla:2013dpa}.
For the $\Upsilon$(1S) a dissociation temperature 
of about 450 MeV is obtained within pQCD~\cite{Laine:2007gj,Escobedo:2008sy}, 
consistent with the lattice results mentioned above. One can provide heuristic arguments 
on how the pNRQCD approach in weak coupling can be generalized to strong 
coupling~\cite{Petreczky:2010tk}. The essential observation in this argument is that 
when the binding energy is small the potential is the same as the energy of static 
$Q\bar Q$ pair and thus can be calculated on the lattice using Wilson loops.

Another line of argument for the existence and definition 
of the static potential beyond weak coupling has been given
in ref.~\cite{Burnier:2012az}. It was shown that if a well-defined 
Lorentzian peak exists in the spectrum of the Wilson loop, 
its late time behavior can be described by a 
Schr\"odinger-like equation with a time independent potential,
whose real and imaginary parts are related to the position and 
width of the peak. Recent works in quenched and full QCD ranging 
up to $\sim2 T_{\rm c}$ have all reported the observation of such well-defined 
peak structures~\cite{Burnier:2014ssa,Bazavov:2014kva,Burnier:2015tda,Burnier:2016mxc}. 
More theoretical work is needed to understand the relation between 
the energy of static $Q\bar Q$ pair at $T > 0$ obtained from Wilson 
loops and the pNRQCD definition of the potential.
 
Using such a static potential and its in-medium modification obtained from the
lattice, the bottomonium and charmonium spectrum in the S and
P-wave channel were computed recently~\cite{Burnier:2015tda,Burnier:2016kqm}
(see \cite{Burnier:2007qm} for a weak coupling analysis). It was found that
the narrow vacuum states are hierarchically modified; this is caused 
by their transitions into open heavy-flavor states or excited bound states
(strong~\cite{Kim:2016zyy} 
or electromagnetic decays are not included). The weakest bound states
are affected most quickly. The modification consists both of a broadening of
the states, as well as a shift to lower masses, which however due to the
concurrent lowering of the open charm threshold still leads to a decrease in
the in-medium binding energy. A comparison of the width and the binding
energy can also be used for defining a 
melting temperature~\cite{Laine:2006ns,Escobedo:2008sy}, which however
does not mean that all features would have completely disappeared from 
the spectral function.

Since on the lattice one cannot decipher the microscopic origin of the 
broadening observed, the width may be related to both the phenomena of
gluo-dissociation and inelastic parton scattering mentioned above, and
to transitions to a color octet state with repelling interaction that
is often discussed within perturbative approaches, which 
in turn contributes to the presence of unbound heavy quarks in the medium.

Apart from reducing systematic errors on the lattice side and going to 
higher order on the perturbative side, an open issue is to address finite
velocity corrections on the lattice. These are expected to be strongly
suppressed in the non-relativistic regime, but nevertheless worth 
a consideration~\cite{Escobedo:2013tca}. \newline 

\noindent{\bf Spectral functions from the Bethe-Salpeter equation}\\
There exists a multi-year effort to also compute the spectra of heavy
quarkonium and open heavy-flavor using the T-matrix 
approach~\cite{Mannarelli:2005pz,Cabrera:2006wh,Riek:2010fk,Riek:2010py}.
In this approach a 3-D reduced version (T-matrix)
of the Bethe-Salpeter equation is used to capture the physics of the
in-medium bound states and/or resonances. In ref.~\cite{Riek:2010py}, 
a step was undertaken toward a self-consistent solution of the 
coupled one-body (self-energy) and two-body (T-matrix) equations 
in the thermal environment, thereby highlighting the role of off-shell 
self-energy feedback to the two-body scattering. The calculated
Euclidean temporal correlator from this work was compared to the
lattice results~\cite{Aarts:2007pk}. The apparent deviation of the calculated correlator
ratio from lattice results at small Euclidean time was due to the fact that
different reference correlators were used in the T-matrix approach
(vacuum reference correlator) and the lattice (reference at a bit
above $T_{\rm c}$).

The concept of a two-body in-medium potential is inherent in this
approach. While the heavy-quark internal energy and free energy from
lattice calculations were used as the input potential in the T-matrix
to bracket theoretical uncertainties in previous works, recently an
in-medium potential has been defined and extracted in this
thermodynamic many-body approach~\cite{Liu:2015ypa}, stipulating in
particular how finite-width effects (in both potential and HQ
propagators) affect the extraction of the underlying interaction
kernel. Yet, it has not been resolved whether or how this potential is
related to the in-medium heavy quark potential discussed above. The
connection of the T-matrix approach to functional methods in QCD, such
as Dyson-Schwinger computations might allow to connect it to the
effective field theory approaches in the future.

The use of Bethe-Salpeter equations has also recently been advocated to
extract a potential for charmonium at finite charm mass~\cite{Iida:2011za,Allton:2015ora}. 
The work applies the HAL-QCD method of extracting a potential for a finite-mass 
two-body system~\cite{Ikeda:2011bs} rigorously defined at zero temperature. 
It has thus raised questions about the applicability of the HAL-QCD method approach 
in the presence of a thermal medium, which are not yet finally settled. \newline

\noindent{\bf Spectral functions in fully relativistic approaches}\\
The methods discussed above can in principle be applied to any type of 
Euclidean correlators. A particular highlight discussed during the workshop
was related to the fate of baryons at finite temperature. 
In particular, first results on parity doubling~\cite{Aarts:2015mma} are 
consistent with expectations: in-medium effects in the
hadronic plasma have been observed, with a stronger effect in the
negative-parity channels. It remains to be seen whether and how this will
impact models based on the hadron resonance gas and statistical hadronisation.

One limitation of the lattice NRQCD approach is the lack of a continuum limit and 
possible limitations of the approximation for charm quarks. For the bottom
quark NRQCD on a full relativistic sea should be accurate, while charm might 
require a fully relativistic lattice QCD approach. Large and fine enough lattices that 
allow for a continuum extrapolation and also allow for a relativistic treatment of bottom 
quarks will become available in the near future but will still be limited to the quenched 
approximation. 

While significant progress has been made the problem of quarkonium properties at
high temperatures is far from being solved. Further improvements in the precision of the
lattice calculations and refinements of the Bayesian methods will be needed, but these alone
will not be sufficient to solve the problem. Clearly, a synergy of different approaches
will be needed to reliably establish quarkonium properties in the high temperature medium.
Examples of such synergy would include the combined use of NRQCD and pNRQCD. One can calculate
the spectral functions in pNRQCD and from these obtain the Euclidean time correlation function
that can be compared to the ones obtained in lattice NRQCD calculations, thus providing a sanity check.
Similarly, the calculation of spatial and temporal correlation functions in the relativistic
approach can be compared to the correlators obtained in the T-matrix approach. Such comparison
could also be performed for open heavy-flavor hadrons. Both pNRQCD and the T-matrix approach would
benefit from improved lattice calculations of the heavy-quark anti-quark potential. These calculations
should be pursued in the future.

\subsubsection{Heavy and light flavor diffusion} \label{ss:diff}
If a light or heavy quark has a momentum of at most of the order 
of the temperature with respect to the medium rest frame, its movement can be 
characterized by a ``transport coefficient'', known as the diffusion
coefficient $D$~\cite{Berrehrah:2016led,Berrehrah:2014tva}. 
The value of $D$ may depend modestly on the quark mass but, 
within the validity of the diffusive description, is independent of 
the momentum (cf.\ e.g.\ ref.~\cite{Hong:2010at} for a review). 

In general, a lattice determination of $D$ is challenging, because
it requires resolving fine features of a spectral function (see above). 
Perturbatively, for heavy quarks, the width of the ``transport peak'' is 
$\sim \alpha_s^2 T^2/m$~\cite{Petreczky:2005nh} and therefore very small. 
However, in broad analogy with the concept of a static potential for
quarkonium physics, an EFT approach permits to reduce also the open 
heavy-flavor problem to a purely gluonic correlator~\cite{CasalderreySolana:2006rq,CaronHuot:2009uh},
which has no transport peak~\cite{Laine:2009dd}. Apart from 
permitting for the first NLO computation of a transport 
coefficient~\cite{CaronHuot:2008uh}, this formulation has led to 
a well manageable lattice challenge~\cite{Banerjee:2011ra}.
By now even the continuum limit has been 
reached~\cite{Francis:2015daa,Christensen:2016wdo}, with a result
that can be directly used in phenomenology.   
Ultimately, the importance of $1/m$ corrections in the case of 
charm also needs to be addressed; these can be reduced to 
a ``higher-order'' gluonic correlator. 

It is important to stress the role played by systematic errors in 
lattice determinations of transport coefficients. The analysis of 
ref.~\cite{Francis:2015daa} contains a realistic estimate of systematic
uncertainties and therefore a large but most likely reliable error bar. 
In contrast, some previous analyses appear to have much smaller 
error bars~\cite{Ding:2012sp}, however the errors are 
statistical only and therefore underestimates. 

For light quarks, the transport peak is broader than for heavy quarks
and therefore the numerical challenge is not quite as hard. Nevertheless,
the perturbative width~\cite{Moore:2006qn} is much smaller than those
indicated by current Bayesian spectral reconstructions, which is 
a reason for further investigating the issue. Indeed, the area
under the transport peak can be well constrained. This implies that if 
a spectral reconstruction overestimates the width of the peak, it 
underestimates its height ($D$). Values of $D$ accounting for 
this uncertainty in the quenched continuum limit 
can be found in refs.~\cite{Ding:2016hua,Ghiglieri:2016tvj}, 
whereas recent results at a finite lattice spacing 
of the unquenched theory can be found 
in refs.~\cite{Aarts:2014nba,Brandt:2015aqk}.

An improved ab-initio calculation of the heavy-quark diffusion constant is needed.
Lattice calculation of this quantity with dynamical quarks is prohibitively expensive
if not impossible. One way to go forward would be to extend the quenched calculations
to higher temperatures and match the weak coupling and lattice results with the help
of a $K$ factor. Assuming that this $K$ factor is independent of the number of quark flavors
one could get an estimate of the heavy-quark diffusion constant for $T > 300$~MeV.

\subsubsection{High-$\pt$ jets} \label{ss:jets}
Heavy-flavor jets generated in the experimental setting have typically
large transverse momenta with respect to the medium. Treating such jets 
on a non-perturbative level poses a conceptual challenge. In phenomenological
studies, a Fokker-Planck equation parametrized by two coefficients, 
$\hat{q}$ and $\hat{q}_{\rm L}$, is frequently used. The basic premise behind
the Fokker-Planck equation is that the number density of the hard particles 
(integrated over momenta $p \gg T$ of the phase space distribution) is 
conserved. However, in QCD there are processes already at leading order 
in $\alpha_{\rm s}$, which violate the Fokker-Planck equation~\cite{Ghiglieri:2015ala}, 
e.g.\ a collinear splitting $Q\to Qg$ where {\em both} the quark and 
the gluon carry a large $\pt$ (``$\pt/2$''), or a scattering of $Q$ with a hard
medium gluon whereby its momentum is changed by a large amount. 

On the lattice, where every process is included in the measurement, it is 
not possible to separate those reactions, which do allow for a Fokker-Planck
treatment from those that do not. 
In contrast, in a perturbative approach, this can be done in 
principle. Consequently, the perturbative side can be ``boosted'' into 
an EFT approach, in which the contributions of certain ``soft'' momentum
scales are resummed to all orders. This has recently led to an approach
in which ``soft contributions'' to the so-called transverse collision
kernel $C(k_\perp)$ can be defined beyond leading order~\cite{CaronHuot:2008ni} 
and subsequently measured through EFT lattice 
simulations~\cite{Laine:2012ht,Benzke:2012sz,Panero:2013pla,Laine:2013apa,D'Onofrio:2014qxa}. 
If the contributions of ``hard scatterings''~\cite{Arnold:2008vd}
are properly added, a value can in principle be obtained for 
$\hat{q}$, as a certain moment of $C(k_\perp)$. 
The value of $\hat{q}_{\rm L}$ remains to be determined.
The un-integrated $C(k_\perp)$ may also parametrize 
a subset of non-Fokker Planck processes~\cite{Ghiglieri:2015ala}. 

Further progress on high-$\pt$ probes can be made through the use
of the EFT approach. In particular, the use of soft collinear effective theory
could be beneficial. The non-perturbative information for the relevant processes
would be then included in the coefficients of this effective theory.

\subsection{Connection to phenomenology and experimental results}

\noindent{\bf Heavy-flavor diffusion and jets}\\
The phenomenological description of diffusion and high-energy jets makes
use of quantities such as the diffusion coefficient $D$ and the jet quenching
parameters $\hat{q}$ and $\hat{q}_{\rm L}$ mentioned above. 
The status of their determination from QCD was discussed in the 
paragraphs above and will not be repeated here. \newline

\noindent{\bf Effective processes}\\
The goal of a model is to simplify the technical description of a particular 
process and at the same time to shed light on the relevant underlying physics. 
Along this way it may borrow concepts from theory even if these are extended 
beyond their formal range of applicability. In the end however, it has to be made 
sure that the model faithfully describes the process it was designed for.

One example is the concept of a screening mass in the context of the in-medium 
modification of the heavy-quark potential. The values of the potential determined 
on the lattice have indeed been reproduced with an Ansatz that combines the 
vacuum physics of a confined $Q\bar{Q}$ in the form of a Cornell potential with 
that of a weakly coupled quark-gluon plasma via the concept of a generalized 
Gauss law~\cite{Burnier:2015nsa}. This leads to analytic expressions for the 
real and imaginary parts of the potential, which depend only on a single temperature 
dependent parameter $m_{\rm D}(T)$ appearing in the form of an in-medium mass. 
While this in-medium mass can now be used to easily describe the in-medium 
modification of the heavy-quark potential it may not be immediately generalized 
for use in other physical situations, such as e.g.\ those including light quarks. 

The concept of effective running coupling 
$\alpha(r, T)$~\cite{Burnier:2016mxc,Kaczmarek:2004gv,Kaczmarek:2005ui,Bazavov:2014soa} 
may be seen in a similar fashion, i.e.\ it represents a quantity that allows a concise
description of processes related to heavy-quark interactions, however its 
value is definition-dependent and may not be used in every context.
It has been applied with success in transport approaches~\cite{Song:2016rzw,Song:2015ykw,Song:2015sfa}.
Another definition of an effective coupling, useful for the EFT description of 
soft observables, can be found in ref.~\cite{Laine:2005ai}. \newline

\noindent{\bf Open quantum systems}\\
In contrast to the classic picture~\cite{Matsui:1986dk}, 
in which quarkonium melts when Debye
screening affects the bound state radius, i.e.\ $r m_D \sim 1$,
it is nowadays believed that the bound state dissolves through dynamical
scattering processes already in the regime $r m_D < 1$, i.e.\ when Debye
screening is not yet efficient~\cite{Escobedo:2008sy,Burnier:2007qm,Dominguez:2008be}.  
In order to describe this dynamics the concept of open quantum systems~\cite{Borghini:2011ms,Borghini:2011yq,Young:2010jq,Akamatsu:2011se,Akamatsu:2012vt,Rothkopf:2013kya,Akamatsu:2014qsa,Akamatsu:2015kaa,Blaizot:2015hya,Brambilla:2016wgg}
has recently received strong interest. 
(Related approaches can be found in, e.g., refs.~\cite{CasalderreySolana:2012av,Krouppa:2015yoa}.)
The separation of scales between the constituent quarks and the thermal medium
invites a natural distinction between environment and subsystem that underlies
this approach. The description can either be based on the time evolution of a
reduced density matrix, i.e.\ a so-called master equation, or on the evolution 
of a particular realization of the subsystem usually involving a wave function.

In the open quantum system approach, 
the real and imaginary parts of the in-medium heavy-quark potential
can be related to the stochastic evolution of the in-medium heavy quarkonium
wave function~\cite{Akamatsu:2011se}. In particular, 
the imaginary part is related to the strength of in-medium noise. This
approach however is only applicable at early times and does not yet include
dissipative effects required for consistent thermalization. Its extension
towards thermalization has been applied to $\Upsilon$ physics 
in a static medium in ref.~\cite{Rothkopf:2013kya}. Currently, 
increasingly advanced theoretical work is underway~\cite{Akamatsu:2015kaa}, 
with the goal of systematically connecting the language of open quantum 
systems with that of effective field theory~\cite{Brambilla:2016wgg}. 
In particular in~\cite{Brambilla:2016wgg}, 
it has been calculated the nuclear modification factor $\raa$ for the states 
$\Upsilon$(1S) and $\Upsilon$(2S) in a strongly coupled plasma.
This is the first calculation that takes into account both the singlet
and octet contributions. The corresponding evolution equations account both
for dissociation and recombination and have promising phenomenological
applications.

At the same time, a new Schr\"odinger-Langevin approach has been put 
forward~\cite{Katz:2015qja}, which, with the inclusion of a non-linear contribution 
in the stochastic evolution of the quarkonium wave function, allows the heavy 
quarkonium state to thermalize at late times. It has been thoroughly explored in 
1-dimensional settings and its formulation in terms of a Schr\"odinger equation 
bodes well for relating its parameters to quantities on the EFT side in the future. 

To improve the phenomenological models by using inputs from lattice QCD a more
detailed understanding of color screening is needed. This can be achieved by comparison
of weak coupling and lattice results on static quark free energy. Further theoretical work
is needed to relate phenomenological models of quarkonia production to the lattice results
on the complex potential. The implications of the octet degrees of freedom in pNRQCD
on the dynamical models of quarkonium production also need to be better understood.

\section{Phenomenology and experiment of open heavy-flavor probes (Discussion Group 2)} \label{sec:dg2}


The Discussion Group envisages three concrete questions:
\begin{itemize}
	\item Is there a tension between the $\raa$ measurements of non-prompt \jpsi{} and B mesons?
	\item Is the pp baseline of open heavy-flavor production under theoretical control? What are the uncertainties in energy-loss predictions and due to the theoretical uncertainties due to the pp baseline?
	\item How can we rule out energy-loss models?
\end{itemize}
In a subsequent dedicated open heavy-flavor discussion, a focus was put on the measurement 
capabilities and community needs related to the future sPHENIX program~\cite{Aidala:2012nz}.
Very valuable bilateral discussions took place with the quarkonia (see Chapter~\ref{sec:dg4}) 
and lattice discussion groups (see Chapter~\ref{sec:dg1}). The general charges for these mutual 
discussions were observables of mutual interest and how to connect relevant quantities in 
energy-loss calculations to quantities that can be computed on the lattice, respectively.

\subsection{Open heavy-flavor production and model descriptions}

\begin{figure}[t!]
\centering
   \includegraphics[width=0.46\textwidth]{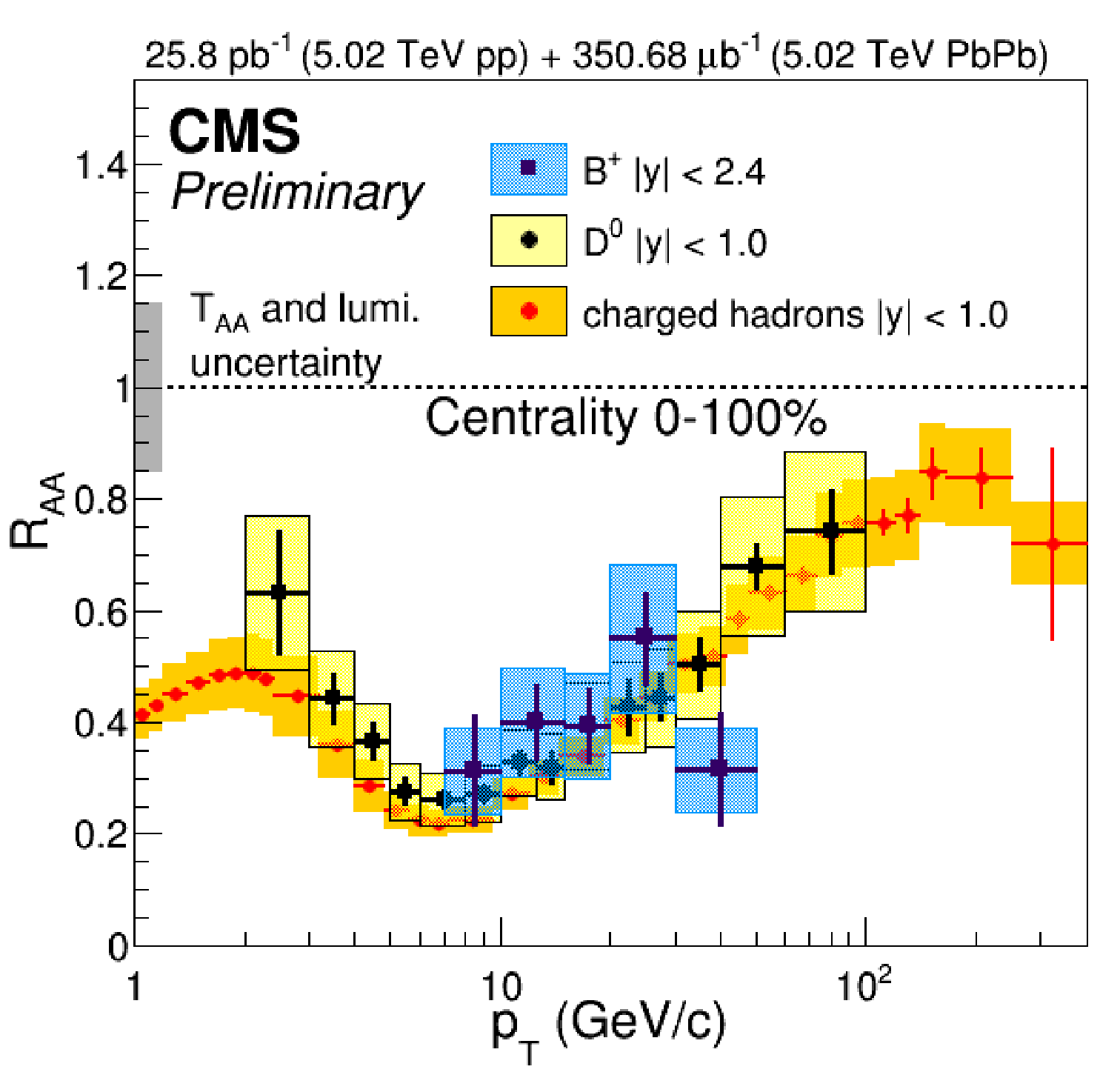}  
\hspace{8mm}
   \includegraphics[width=0.42\textwidth]{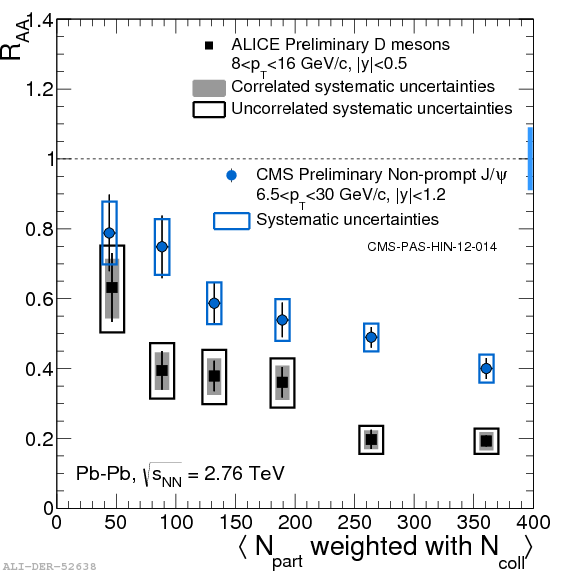}
\caption{(Top) Nuclear modification factor as a function of $\pt$ for B mesons \protect\cite{CMS:2016jya}, D mesons \protect\cite{CMS:2016nrh}, and non-identified charged hadrons \protect\cite{CMS:2016ouo} in centrality integrated Pb-Pb collisions at $\snn~=$ 5.02~TeV. (Bottom) Nuclear modification factor as a function of the number of participants for non-prompt \jpsi{} and D mesons in Pb-Pb collisions at $\snn~=$ 2.76~TeV \protect\cite{Jena:2014xga}.}
\label{fig:BDh}
\end{figure}
\begin{figure}[t!]
\begin{center}
\includegraphics[width=0.44\textwidth]{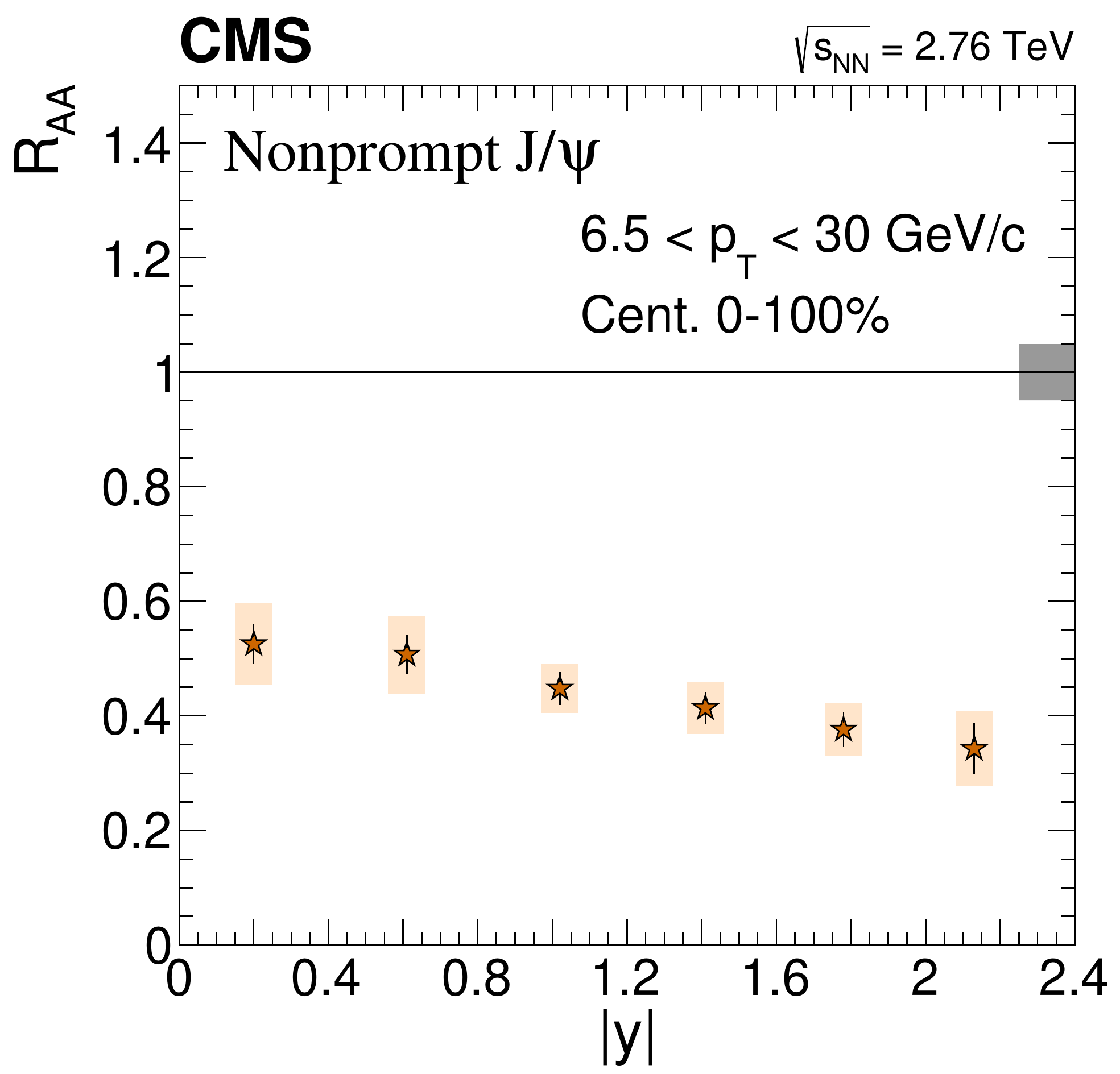}
\hspace{4mm}
\caption{Nuclear modification factor $\raa$ as a function of rapidity $y$ for non-prompt \jpsi{} in Pb-Pb collisions at $\snn~= 2.76$~TeV \protect\cite{Khachatryan:2016ypw}.}
\label{fig:Jpsiy}
\end{center}
\end{figure}

\subsubsection{B mesons}
At the Hard Probes conference 2016 in Wuhan, China, the CMS experiment showed for the 
first time the nuclear modification factor $\raa(\pt)$ for open B mesons~\cite{CMS:2016jya}. As seen in the upper panel of 
Fig.~\ref{fig:BDh}, the measured CMS suppression of the open B mesons is the same as the 
D mesons~\cite{CMS:2016nrh} and charged hadrons in the measured $\pt$ range~\cite{CMS:2016ouo}, also measured by CMS. 
These results agree with ALICE measurements of the D-meson and charged hadron 
$\raa$~\cite{ALICE:2012ab,Adam:2015sza,Abelev:2012hxa}. 
This consistency of suppression across parton flavors was a surprise on both theoretical 
and experimental grounds.  
From the theoretical side, generically all energy-loss models predict less B-meson suppression than for D mesons and charged hadrons. 
At the partonic level, energy loss decreases as a function of the parton mass due to the suppression of small angel gluon radiation \cite{Dokshitzer:2001zm}. Predictions for the suppression of hadronic observables convolve the different production spectra of the various parton flavors with their energy loss and then subsequent relevant non-perturbative fragmentation functions. As a result, there can be a non-trivial momentum dependence to the hadronic flavor ordering predictions as a function of momentum, with D mesons generally as suppressed as charged hadrons at low $\pt$~\cite{Buzzatti:2011vt,Horowitz:2012cf,Djordjevic:2013xoa}.  

Experimentally, a well-known \cite{Jena:2014xga} combination of \\ $\raa(N_{\rm part})$ of non-prompt \jpsi{}, the decay products of B mesons, measured by CMS~\cite{Khachatryan:2016ypw} and D mesons by \mbox{ALICE} \cite{ALICE:2012ab,Adam:2015sza} in a single plot showed a clear separation of heavy-flavor hadrons by mass, as can be seen in the lower panel of Fig.~\ref{fig:BDh}.  

The rapidity selection of the data was identified as playing an important role. CMS measured charged hadrons and D mesons in $|y| < 1$ as compared to $|y| < 2.4$ for the B mesons. Previous CMS results~\cite{Khachatryan:2016ypw} show that the non-prompt \jpsi{} $\raa(y)$ decreases surprisingly quickly with increasing $|y|$, as depicted in Fig.~\ref{fig:Jpsiy}. It is worth noting that 1.) there is a CMS non-prompt \jpsi{} $\raa(\pt)$ measurement  with the same $|y| < 2.4$ rapidity cut as for the B mesons, and these two measurements show a consistent suppression and 2.) there is no immediately obvious theoretical explanation for the decrease in \jpsi{} $\raa(y)$.

The members of the CMS collaboration present committed to measuring the spectra with the same 
rapidity selections, and ultimately differentially in $\eta$ and $\pt$, with future larger data samples. 

\subsubsection{pp baseline}
\begin{figure}[t!]
\centering
   \includegraphics[width=0.44\textwidth]{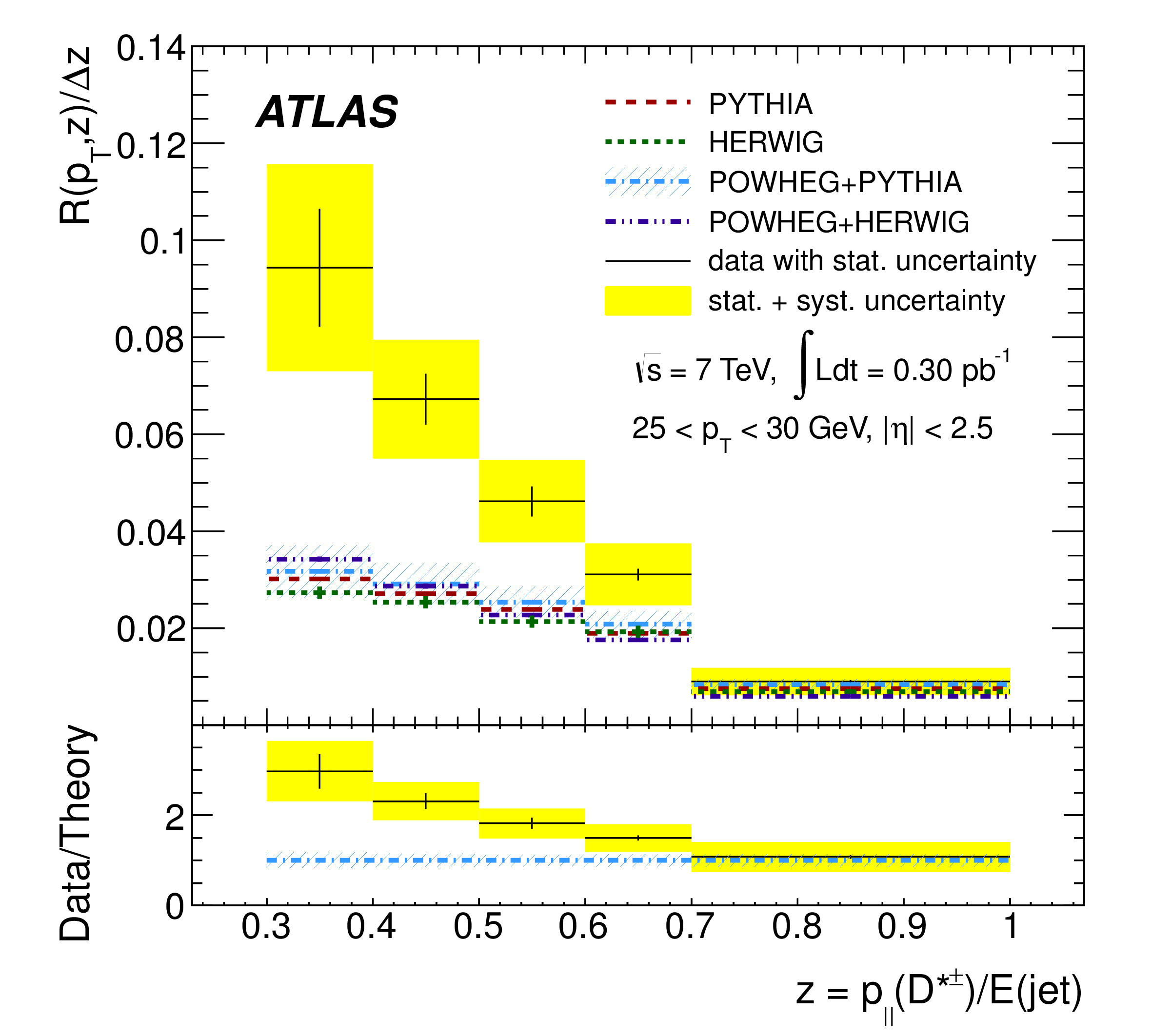}
\hspace{8mm}
   \includegraphics[width=0.42\textwidth]{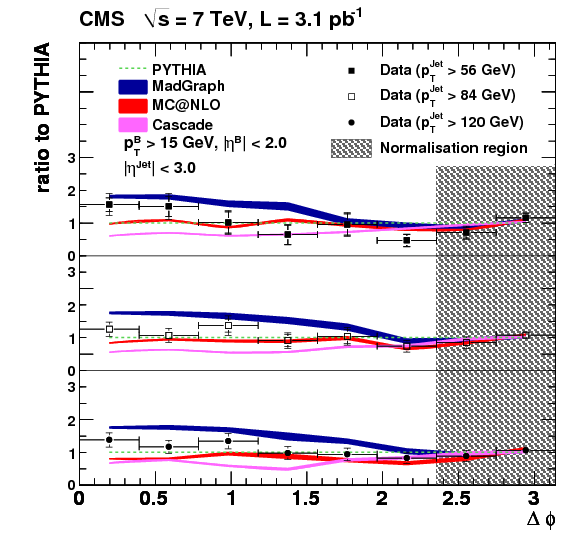}  
\vspace{-2mm}
\caption{(Top) Comparison of the D$^{*\pm}$ production rate measured in ATLAS and several Monte Carlo generators \protect\cite{Aad:2011td}.
(Bottom) Azimuthal angular correlation distribution between B$\overline{\rm B}$ pairs in pp collisions at $\sqrts~=$ 7~TeV measured by CMS. The data are compared to several generators \protect\cite{Khachatryan:2011wq}.}
\label{fig:bbar}
\end{figure}

Through discussions at this workshop, there was a general appreciation for the significant theoretical uncertainties associated with single open heavy-flavor production via fixed order at next-to-leading-log (FONLL)~\cite{Cacciari:2001td} or the general-mass variable flavor number scheme (GM-VFNS) \cite{Kniehl:2004fy,Kniehl:2005mk} and agreement that future energy-loss calculations should explore the propagation of uncertainties due to the production calculations through to their final suppression predictions.  Fragmentation functions are an additional concern: it seems that both light and heavy-flavor fragmentation functions fail to describe top energy LHC measurements~\cite{Aad:2011td,d'Enterria:2013vba}; see the upper panel of Fig.~\ref{fig:bbar}, which shows the differential production rate of $D^{*}$ mesons (per-jet) as a function of $z$.  There is also now evidence for double parton scattering in heavy-flavor production, which is not included in any generator \cite{Aaij:2012dz,Aaij:2015wpa}. The community will need to seek advice from the heavy-flavor production practitioners to understand in detail the extent to which the $c$ and $b$ spectra can be varied to accurately reflect the uncertainties in the shape of the production spectra.

There is a general consensus amongst heavy-ion experimentalists that state-of-the-art heavy-flavor generators do not reproduce correlations measurements in pp collisions.  Through discussions at the workshop, it was realized there are extremely few measurements related to the correlations of heavy flavor, in either angle or in momentum.  It is also clear that those measurements that do exist are not necessarily well described by current state-of-the-art NLO heavy-flavor generators, particularly for collinear production of heavy-quark pairs~\cite{Khachatryan:2011wq,Chatrchyan:2013zja}; see the lower panel of Fig.~\ref{fig:bbar}.  
Nevertheless, there was consensus amongst the participants that 1.) the future of the heavy-flavor community lies in distinguishing measurements related to correlated observables and 2.) the self-consistent merging of sophisticated NLO production codes (and their matching to parton showers and hadronization) with energy-loss calculations is an open problem.

\subsubsection{Model tests}
In the last years, several groups have advanced approaches, which study the interaction of heavy quarks in an expanding QGP~\cite{Berrehrah:2014tva,Song:2016rzw,Song:2015ykw,Song:2015sfa,Gossiaux:2008jv,Aichelin:2013mra,Nahrgang:2016lst,Uphoff:2014hza,Uphoff:2012gb,Das:2013kea,Das:2016cwd,Das:2015ana,vanHees:2007me,He:2014cla,He:2012df,He:2011qa,Cao:2013ita,Cao:2015hia,Cao:2016gvr}.
These approaches are either based on a Fokker-Planck equation or use the full Boltzmann collision kernel to describe the collisions of heavy quarks with the partons of the QGP. Also different approaches for the plasma expansion are used. For bottom quarks the Fokker Planck and the Boltzmann approach yield almost identical results. For the charm quarks differences between both approaches have been reported~\cite{Das:2013kea} what should be investigated in detail in near future.

All these approaches come to the common conclusion that the standard perturbative QCD cross sections~\cite{Moore:2004tg} are not sufficient to describe the observed energy loss of heavy quarks in the plasma~\cite{Berrehrah:2014tva,Berrehrah:2014kba}.
The model differ in the way in which non-perturbative elements are included. Some of them have modified the Debye mass to have a smooth transition between collisions with low momentum transfer, which require hard thermal loop calculation and collisions with large momentum transfer, which are described by pQCD
~\cite{Gossiaux:2008jv,Aichelin:2013mra,Nahrgang:2016lst,Uphoff:2014hza,Uphoff:2012gb}. 
Others use temperature dependent coupling constants~\cite{Berrehrah:2014tva,Song:2016rzw,Song:2015ykw,Song:2015sfa,Das:2013kea,Das:2016cwd,Das:2015ana}, which increase towards the critical temperature $T_c$. A third approach is based on the existence of heavy quark - light quark bound states close to $T_c$~\cite{vanHees:2007me,He:2014cla,He:2012df,He:2011qa}.

The observables studied in these approaches are the $\raa$ and $v_2$. For heavy quarks with low transverse momentum the predicted $v_2$ values are sensitive to the late stage of the expansion because the heavy quarks get their elliptic flow by collisions with the light plasma partons. 
This is only possible after the spatial eccentricity of the plasma constituents is converted into $v_2$ in momentum space. This takes time. The $v_2$ at higher transverse momenta is caused by different path lengths of the heavy quarks in the QGP~\cite{Nahrgang:2014vza}. Whereas the temperature dependent coupling constant as well as the existence of resonances provoke a high collision rate close to $T_c$, where the elliptic flow of the light partons is already developed, the modified Debye mass increases the collision rate during the whole expansion process. In principle, with more precise data, one may therefore hope that experimental data can decide which is the more realistic scenario.   

It was suggested that moderate $\pt\sim 10$ GeV/$c$ is the ideal momentum region to experimentally test the predicted flavor hierarchy of $b$-decay products compared to $c$ and $u/d/s/g$ decay products because $m_{\rm B}/\pt$ is large and the region will have a good overlap of experimental observables with sufficient statistics at RHIC and LHC.  
Thus, precise measurements of $\raa(\pt)$ for B are considered the highest priority in the field. At the same time, both heavy-flavor production and many energy-loss calculations assume $\pt \gg m_{\rm Q}$ and fragmentation functions appear to be under the best control for $\pt\gg m_{\rm Q}$ (as inferred from the anomalous baryon to meson ratio), and so the most phenomenologically relevant momentum region is currently under the least theoretical control.

In general, there was consensus that correlation measurements, especially in momentum \cite{Tsiledakis:2009da,Gossiaux:2014WWND,Hambrock:2016SAIP}, hold great promise for providing a distinguishing future measurement. However, there is currently a lack of distinguishing predictions from a wide range of theoretical energy-loss models.

While not mentioned at the workshop, comparing measurements across the $\snn$ lever arm, with its attendant change in temperature profiles as a function of time, should provide experimental constraint on energy-loss models. Additionally, although their specific purpose is not ``model killing'', there are two ongoing collective actions begun in 2016 -- the {\em Heavy Quark Working Group}~\cite{HQWG} and the {\em EMMI Rapid Reaction Task Force}~\cite{EMMIRRTF} -- devoted to the extraction of transport coefficients by confronting theoretical models with experimental data and to investigate the different physics behind the various approaches. Such collaborative work will necessarily lead to some standardization, as was the case for the JET Collaboration \cite{JET,Burke:2013yra}, and will point towards improvements needed for some of the models.


\subsection{Discussion on differential and precision measurements}
\begin{figure}[t!]
\begin{center}
\includegraphics[width=0.5\textwidth]{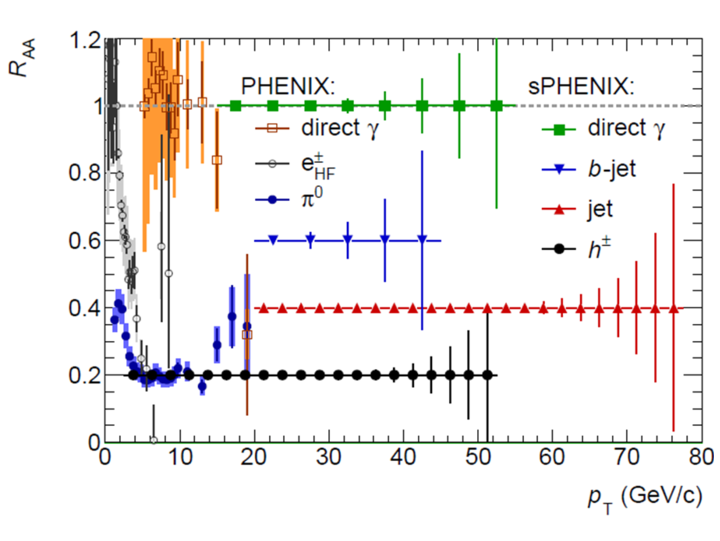}
\vspace{-.2in}
\caption{Projections for measurements of $\raa$ for various observables with the sPHENIX detector~\cite{Aidala:2012nz}.}
\label{fig:sPHENIX}
\end{center}
\end{figure}

There was consensus that theoretical calculations must describe data across flavor, $\sqrts$, $\pt$ and centrality dependencies. Thus, future sPHENIX bottom related measurements will provide a critical cross check of our physical understanding of quark-gluon plasma formation in colliders. With the current projection for the detector setup, which does not include the possibility of particle identification, measurements at sPHENIX will be limited to $b$-jets, charged hadrons, and, possibly, to non-prompt \jpsi{}, as shown in Fig.~\ref{fig:sPHENIX}~\cite{Aidala:2012nz}; there will be no $c$ quark related measurements. 
It was pointed out, however, that it may be possible to design the time projection chamber to provide some particle identification (PID), or even add a dedicated time-of-flight detector. These changes would allow measurements of prompt and non-prompt \jpsi{} and D mesons, 
D$\overline{\rm D}$ correlations, and possibly even (exclusive) B mesons.

There was consensus amongst the participants that measure\-ments in what was termed the ``low-$\pt$'' $\lesssim15$ GeV/$c$ region are the most interesting from a theoretical viewpoint. At these low momentum scales, the bottom-quark mass is similar to its momentum. One expects from perturbative physics a transition from radiation dominated energy loss to collisional energy loss for bottom quarks~\cite{Aichelin:2013mra,Cao:2015hia,Wicks:2005gt}; 
at higher momenta the pQCD-based calculations are expected to be under better theoretical control due to asymptotic freedom. From the AdS/CFT approach, the energy-loss calculations are under the most control at the lowest momenta: momentum fluctuations calculations~\cite{Gubser:2006nz,CasalderreySolana:2007qw,Akamatsu:2008ge,Moerman:2016wpv} agree exactly at $p=0$ but then differ by various powers of the Lorentz boost $\gamma$~\footnote{Since the strength of momentum fluctuations from some of these calculations~\cite{Gubser:2006nz,CasalderreySolana:2007qw} do not satisfy the fluctuation-dissipation theorem, there is some debate as to whether the origin of the momentum fluctuations in those calculations is in fact thermal or is an artefact of the calculational setup \cite{Moerman:2016wpv}.}. 
From an experimental standpoint, this momentum region is of significant interest as there will be overlap of statistically significant measurable quantities at both RHIC and LHC. 

\subsection{Energy loss and quarkonia joint discussion}
A discussion of potential common ground between the two discussion groups began with the topic of boosted quarkonia, where the question was posed ``In what kinematic range (if any), does energy loss become the dominant mechanism of nuclear modification for quarkonia?'' This question arises naturally based on the flatness of $\raa$ at large values of $\pt$ measured for prompt \jpsi{} and $\Upsilon$(1S), with a value similar to that of other hadron species. It was pointed out that the relevance of color singlet versus octet configurations in quarkonium production at large $\pt$, which is still not definitively understood, is crucial to address this question. On the other hand, a comparison with heavy-quark pairs in the antenna configuration 
(i.e.\ propagating with small opening angle), e.g., as measured with ``fat'' jets with two b-tags, could be informative.

The need to further understand quarkonium production led to a discussion of measurements of associated hadro-production. The use of quarkonia with two particle correlations, e.g., \jpsi{}-hadron, and/or jet reconstruction, e.g., to measure the \jpsi{} jet fragmentation function, were pointed out as promising directions. With larger data samples, photon+quarkonium and double quarkonium production are also of interest, as these quarkonia should then initially be produced in the octet configuration, although care has to be taken to understand the contribution of double parton scattering. 

The diffusion of the open heavy flavor prior to combination into quarkonia and quarkonia break-up due to momentum fluctuations from the medium that can be related to the diffusion coefficient must affect the low-$\pt$ production of quarkonia. Thus, measurements of quarkonia down to $\pt = 0$ might provide a complimentary way to constrain experimentally the heavy-quark diffusion coefficient.

Finally, the issue of heavy-flavor correlations, such as D$\overline{\rm D}$, was raised. Such correlations are considered a promising avenue to understand heavy-quark energy loss~\cite{Nahrgang:2013saa,Cao:2015cba,Mischke:2008af}. From the point of view of dilepton resonances and the dilepton continuum, such heavy-flavor correlations can be considered a background to other processes, e.g., Drell-Yan production. The question was posed to what extent the interest in such measurements is convergent between the two groups. In particular, is there interest in low $\pt$ and/or low invariant mass measurements from the energy-loss point of view. The response to this question was found to be model dependent, with different models that incorporate energy loss being interested in different kinematic regimes. This was pointed out as an area that needs further elucidation from the theory side. In particular, it would be desirable to compare predictions for heavy-flavor correlations with different models using the same kinematic selections. 

\subsection{Energy loss and lattice joint discussion}
There was significant interest from both the energy loss and lattice communities to make contact between the areas of research. In particular, there is a strong desire from the energy-loss community for guidance from the lattice community on the temperature and momentum dependence of several quantities of direct interest in energy-loss models, such as the heavy-quark diffusion coefficient, the Debye screening scale, whether heavy-light resonances persist for $T > T_{\rm c}$, the running of the coupling with temperature (albeit this is an observable dependent desire) and $\hat{q}$. The lattice community emphasized that the quantities they can compute must be gauge invariant objects related to imaginary time correlators.

\subsubsection{Heavy-flavor diffusion coefficient}
Much of the discussion focused on the calculation of the heavy-flavor momentum diffusion coefficient $D$. 
In principle, $D$ is a function of temperature $T$ and heavy-flavor mass $m_Q$. One may also consider extending possible definitions of $D$ to larger momentum values, outside the hydrodynamic regime.
There are currently two calculations of $D$ from quenched lattice QCD: one at finite heavy-quark mass (but without a continuum extrapolation)~\cite{Ding:2012sp} and one with an infinite heavy-quark mass (but with continuum extrapolation)~\cite{Francis:2015daa}. The results from the two approaches are currently consistent within the large systematic uncertainties. Both of these lattice calculations are currently limited to quenched approximations and require $p = 0$.  
Going beyond the quenched approximation will be difficult numerically but not conceptually.  
Going beyond the small momentum limit is conceptually difficult. A number of suggestions were made:
\begin{itemize}
	\item List the variety of relevant transport coefficients, e.g.\ the drag coefficient and the longitudinal and transverse diffusion coefficients, and determine gauge invariant means of making contact with these quantities through lattice calculations.
	\item Examine the $T$ and $p$ dependence of the pole structure of the spectral function of the heavy quark current at asymptotically high temperatures via pQCD techniques.  
	When weak coupling calculations of the spectral functions for arbitrary spatial momenta are available one can attempt to make contact with lattice QCD studies of the current correlators at different spatial momenta.
	\item Perform a Taylor expansion in $p$ of the relevant four point function computed on the lattice. Such a calculation can be done on the lattice, 
	although it remains technically very difficult to compute spectral functions, also at $p\neq 0$. Note that this would include
	the contributions from both above and below the light cone, and include the correct smearing effects.
	\item Take the phenomenologically extracted $D(T, m_{\rm Q}, \pt)$ from energy-loss model comparisons to data, derive a related spectral function, and compare to pQCD and lattice calculations.
\end{itemize}

\subsubsection{Debye screening}

Additional discussion focused on lattice calculations of several of the dynamical masses commonly used in energy-loss calculations \cite{Armesto:2011ht,Djordjevic:2011dd}: the Debye electric screening mass $\mu_{\rm E}$ and the magnetic screening mass $\mu_{\rm M}$ defined in terms of gluon propagator.  
Lattice calculations of these masses depend on the choice of the gauge~\cite{Karsch:1998tx,Cucchieri:2000cy,Cucchieri:2001tw}, though in some class of gauges the gauge parameter dependence is mild~\cite{Nakamura:2003pu}. At leading order the electric mass is gauge independent. 
A gauge invariant definition of the electric and magnetic screening masses is possible on the lattice (see e.g. Ref.~\cite{Maezawa:2010vj}), but it is not clear how these definitions are related to the finite temperature gluon propagator used in the energy-loss calculations.
A potentially useful approach is to use lattice calculations to measure the transverse momentum squared per unit distance imparted to a high momentum parton, $\hat{q}$ \cite{Panero:2013pla}. 


\subsubsection{Other points of contact}
Additional points of contact between energy loss and lattice calculations include computing the polarization loop or quark number fluctuations from the energy-loss side and comparing to lattice calculations; the susceptibility is another potential observable mutually calculable from the energy-loss perspective and from the lattice.







\section{Phenomenology and experiment of quarkonium production (Discussion Group 3)} \label{sec:dg4}

\subsection{Open experimental and theoretical issues}
LHC measurements are now reaching a high level of accuracy, thanks in particular to the Run2 data taking where a large luminosity has been collected and a considerable effort has been done to reduce the systematic uncertainties. The comparison of the experimental results with the theory calculations is, however, still limited by the large uncertainties on the model inputs. The nuclear modification of the parton distribution functions (the so-called ``shadowing'') has very large uncertainties, since very little experimental constraints are available, in particular for gluon distributions. At present, the charm cross section is the other main source of uncertainty, and also in this case limited experimental guidance is available. The usage of very different input values complicates even further the comparison between data and theory models~\cite{hfGeneral} and an agreement on the values to be adopted, possibly driven by the available experimental results, would clearly ease such a comparison.
The importance of the experimental measurement of the charm cross section, both at forward and mid-rapidity, i.e.\ in the kinematic range covered by the experiments, is, therefore, considered as fundamental to gain further insight in this field. The cross sections should be directly measured in pA and AA collisions, to simultaneously take into account the charm production cross section and its medium modifications. The precision on such measurements, required to provide a meaningful comparison between data and experiments, should be smaller than 10\%. Given the role played by bottomonium ($\Upsilon$) states, in particular at the LHC and in the forthcoming sPHENIX experiment at RHIC, a similar cross section measurement for open-bottom is also mandatory.

The quarkonium production in \mbox{pA} and \mbox{AA} collisions is affected by the so-called {\em cold nuclear matter} (CNM) effects. While these mechanisms are dominant in \mbox{pA} interactions, an extrapolation is needed to estimate their role in \mbox{AA} collisions, where CNM effects are underlying the QGP-related ones.
Advantages and limitations of an approximation for CNM effects in Pb-Pb at forward rapidities, based on the factorization of CNM effects measured at forward and backward rapidities, 
in \mbox{pA} collisions, have been debated. While it is clear that this approach provides at first order the size of some of the currently expected CNM mechanisms in 
\mbox{AA}, it is still a model dependent definition and there are several assumptions, which may limit its validity. In case of shadowing, as dominant CNM mechanism, the 
\mbox{pA} data should cover the same $x$ range as in \mbox{AA}, while in case of coherent energy loss, \mbox{pA} and \mbox{AA} data should be compared at the same center of 
mass energy. Factorization holds also if the $c\bar{c}$ pair absorption in the nucleus is the main CNM effect, provided that a unique absorption cross section describes the 
rapidity dependence of the quarkonium production. However, this situation is realized only at energies much lower than available at the LHC. The validity of the factorization approach has to be assessed, for both $\pt$-integrated and $\pt$-differential results, in all the theory models describing the quarkonium results in \mbox{pA} interactions.

Finally, the current interpretation of the medium modification of the quarkonium states is still limited by the absence of precise feed-down measurements. Recent LHCb results point to a maximum feed-down contribution to the $\Upsilon$(1S) state, from the $\chi_{\rm b}$, $\Upsilon$(2S) and $\Upsilon$(3S) states, of the order of 30\%~\cite{Aaij:2012se}. 
However, the present measurements do not cover the kinematic range of the \mbox{AA} bottomonium results and precise measurements at the LHC, in particular for the $\chi_{\rm b}$, extended down to zero $\pt$ would help to precisely assess the feed-down contribution. For the interpretation of the $J/\psi$ results, the feed-down from the $\chi_{\rm c}$ is an experimentally challenging key-measurement, to be performed in the near future. However, also in this case, the extension of the $\pt$ coverage down to zero turns out to be crucial for this measurement.

\subsection{New observables}
To address these open experimental and theoretical issues, possible new
observables have been discussed at the workshop and are briefly reviewed in the
following.

\subsubsection{Quarkonia}
The nuclear modification factor $\raa$ is the most widely used observable for quarkonium studies both by RHIC and LHC experiments. However, the information conveyed by the $\raa$ is strongly bound to the evaluation of the \mbox{pp} reference. From the experimental point of view, the often limited statistics of the \mbox{pp} reference, or even the absence of \mbox{pp} data collected at the same center of mass energy as \mbox{AA} collisions, might result in a limitation to the accuracy achievable in the $\raa$. From the theory side, the use of \mbox{pp} collisions as a reference holds only under the assumption that quarkonia are formed in inelastic scatterings before QGP is formed. If this is not the case, the relevant reference should be the total charm or bottom production cross section per unit of rapidity. Clearly the two approaches are equivalent as long as the open heavy-flavor cross section scales with the number of binary collisions.

The importance of new quarkonium observables to complement the information provided by the $\raa$ measurement, in \mbox{pA} and \mbox{AA} collisions, has been discussed. It was agreed that experiments should also provide quarkonium yields in the colliding systems under study, to be compared with similar quantities extracted from theory calculations. The quarkonium elliptic flow ($v_2$) and the ratio of the average quarkonium transverse momentum square in \mbox{AA} and \mbox{pp} collisions ($r_{\rm aa}$) can also provide additional informations and theory models should address all these aspects in a consistent way. The quarkonium polarization is considered an interesting observable, even if experimentally the measurement is challenging and, at present, no theory guidance on the expected degree of polarization in \mbox{pA} and \mbox{AA} collisions is provided.
When studying excited and ground quarkonium states, the ratio of the two yields, even normalized to the corresponding yields in pp collisions, might provide additional insight, in particular if a partial cancellation of the experimental and theoretical uncertainties is expected. However, to provide a full picture of the fate of quarkonium resonances in \mbox{pA} or \mbox{AA} collisions, it has been stressed that results provided in terms of ratios should always be accompanied by quarkonium yields and $\raa$ measurements.

The importance of the quarkonium normalization to the open heavy-flavor production is well assessed and the feasibility of such a study has been extensively addressed. To investigate the centrality dependence of the quarkonium production, the open heavy-flavor production has to be experimentally measured down to zero transverse momentum ($\pt$), with an accuracy that should be smaller than 10\%, in order not the represent a limit for the precision of this observable. For $\pt$-differential studies theory guidance is crucial to link the quarkonium and the charm/bottom transverse momentum. The proposed normalization should be investigated both for charmonium and bottomonium resonances, comparing the yields to the open charm and bottom mesons, respectively.

RHIC and LHC quarkonium $\raa$ results are usually presented as a function of the number of participant nucleons ($N_{\rm part}$), evaluated within a Glauber model. 
While this variable represents an easy way to compare to theory calculations, 
 the use of additional variables as the number of charged produced particles in a given rapidity range (d$N_{\rm charged}$/d$\eta$), possibly normalized to the 
 transverse area, is proposed. These quantities, whose precise definition should be agreed among the experiments, are more correlated to the energy density reached in the collisions, thought to be responsible for the underlying physics.

\subsubsection{Open heavy-flavor correlations via dileptons}
The dilepton continuum between the $\phi$ and $J/\psi$ masses is dominated by
simultaneous semileptonic decays of correlated $\mathrm{D}\overline{\mathrm{D}}$
mesons. The shape of the invariant mass spectrum is sensitive to the initial
angular correlation of the $c\bar{c}$ quark pair~\cite{Adare:2009qk,Bedjidian:2004gd}. 
To leading order these are produced back-to-back and lead to relatively high-mass 
dileptons with low pair-$\pt$. Gluon-splitting on the other hand creates dileptons 
with small invariant mass and high pair-$\pt$. A double differential study in pp 
collisions should be able to constrain the relative importance of various production 
mechanisms. In heavy-ion collisions, heavy-quark energy loss will lead to a 
modification of the dilepton invariant mass spectrum. Besides a softening due 
to high-$\pt$ suppression, the invariant mass spectrum is also sensitive to an 
angular decorrelation. In contrast to D mesons, in decays of heavier B mesons 
any initial correlation is washed out. The overall sensitivity to various
energy-loss mechanisms, discussed in Chapter~\ref{sec:dg2}, still needs to be 
studied.

\subsection{Open issues on theory and phenomenology aspects}
During the meeting, some discussion was convened together with the other Discussion Groups. 
The main points are summarized in the following.

\subsubsection{Spectral functions from lattice QCD versus experiment}
As discussed in Chapter~\ref{sec:dg1} (section A.2.), lattice QCD calculations show strong in-medium modifications of quarkonium spectral functions. In particular, a strong broadening around the pole mass is observed~\cite{Brambilla:2013dpa}. With a width of about 200 MeV, the quarkonium lifetime becomes short enough to decay within the QGP and, hence, the broadening should be observable in the dilepton decay channel, akin the broadening of the $\rho$ meson. At present, there is no experimental evidence for a deviation from the vacuum line shape. However, there has been no real effort to look for such modifications that might hide below the radiative tail. Given that one has not observed a modification of the $\phi$ meson, having a vacuum width between the $\rho$ and the $J/\psi$ or $\Upsilon$(1S), a quantitative estimate for the yield of in-medium quarkonium decays would be important before an experimental search. The situation for the $\Upsilon$(1S) might be better than for the $\phi$, as it can exist (and decay) also in the QGP phase, while the $\phi$ feels only the relatively short hadronic gas phase. Lattice QCD, formulated in Euclidean time, cannot provide these estimates so phenomenological models are required. 

If the quarkonia would be in equilibrium with the me\-dium, then the measurement of the broadening would be equivalent 
to the measurement of the suppression, as the probability to decay into dimuon pairs equals 
$\Gamma_{\mu^+\mu^-}/\Gamma_{\rm Tot}(T)$ with $\Gamma_{\mu^+\mu^-}$ 
being the partial width and $\Gamma_{\rm Tot}(T)$ the total width. 
The probability to decay as a dimuon pair would then be 
$\int_{0}^{\infty} \Gamma_{\mu^+\mu^-} e^{-\Gamma_{\rm Tot}(T(t)) t} dt$ while the average width for the dimuon pair would be
$$ \bar{\Gamma}_{\mu^+\mu^-} =\frac{\int_{0}^{\infty} \Gamma_{\rm Tot}(T(t))
  e^{-\Gamma_{\rm Tot}(T(t)) t} dt}{\int_{0}^{\infty} e^{-\Gamma_{\rm Tot}(T(t)) t} dt}.$$

This encompasses both the case of small broadening ($\Gamma_{\rm Tot}\approx \Gamma_{\mu^+\mu^-}$), 
in which case \mbox{$\bar{\Gamma}_{\mu^+\mu^-} \approx \Gamma_{\mu^+\mu^-}$} 
and the case of finite broadening. One should make a detailed study of the integral 
in the numerator to see if a small time/temperature window contribute to the broadening.
However, the problem is that the relative probability of quarkonia decaying into 
dimuons versus total decay probability is several orders of magnitude smaller. 
Even if it would be extremely interesting to experimentally detect a broadening 
despite such small branching ratio, a quantitative understanding would require 
understanding the vacuum branching ratio of quarkonia into dileptons at much 
higher precision than available today.

\subsubsection{Phenomenological model descriptions of quarkonium formation}
In this section, the phenomenology of the model descriptions of quarkonium formation is discussed. \\
 
\paragraph{Quarkonium formation time\\}
An important debate was triggered on the time it takes to form a particular quarkonium state, the so-called {\it formation time} $t_{\rm F}$. Although it cannot be defined rigorously outside a specific model and a specific environment, this concept is commonly used in order to guide the phenomenology, both for pA and AA collisions. A rough estimate of the
formation time can be achieved based on semi-classical considerations: assuming a state can only be well defined after the two heavy quarks have rotated at least once around each other leads to $t_{\rm F}\approx 1/(m_Q v^2)$ for Coulomb states where $v$ is the velocity of the heavy quark in the bound state. Previous work relying on dispersion relations~\cite{Kharzeev:1999bh} came with an estimate $t_{\rm F}\sim 1/(m_2-m_1)$ with ground state mass $m_i$, close to the Heisenberg time, of the order of $0.44~{\rm fm}/c$ for $J/\psi$ and $0.32~{\rm fm}/c$ for $\Upsilon$.

Historically, the equilibration times $\tau_0$ in heavy-ion collisions  were believed to be much larger than the quarkonium formation times, as relying on dynamics involving softer scales. However, the estimation of these equilibration times have shrunk over the past decade~\cite{Gelis:2013rba,Teaney:2009qa,Shuryak:1992wc}, challenging this assumption. 

During the workshop, no general consensus on whether quarkonia enter the QGP as a fully formed bound object was reached. However, this is not really important for the understanding of nuclear modification factor as will be discussed below. The basic point is that dissociation of quarkonia does not necessarily imply that the heavy $Q$ and $\bar Q$ pairs resulting from the dissociated quarkonia become totally uncorrelated. As long as the correlation persists there is a chance that at least some of the $Q$ and $\bar Q$ pairs will form a quarkonium state again. So, one has to deal with the formation of quarkonium states inside the QGP, irrespective whether the formation time of quarkonium is smaller or larger than the QGP formation time. \\

\paragraph{Models of quarkonium formation in heavy-ion collisions\\}
There have been many attempts to explain the nuclear modification factor of $J/\psi$ in terms
of sequential suppression picture (see e.g.~\cite{Karsch:1987pv,Digal:2001ue,Karsch:2005nk}).
However, it was realized already in the early 2000s that (re)generation of charmonia inside the
QGP is possible and will affect the $J/\psi$ nuclear modification factor significantly
(\cite{Thews:2000rj,BraunMunzinger:2000px,Grandchamp:2001pf,Zhou:2014kka,Andronic:2016nof} and references therein). The models based on these ideas were able to explain the $\raa$ of $J/\psi$ at RHIC and 
successfully predicted the $J/\psi$ $\raa$ at LHC (see e.g.\ ref.~\cite{Zhao:2011cv}). 
Models based on the sequential suppression picture (see e.g.\ Refs. \cite{Strickland:2011aa,Krouppa:2015yoa}), 
as well as models that include in-medium bottomonium formation, fairly describe the observed pattern of bottomonium
production in heavy-ion collisions at RHIC and LHC. 
However, a quantitative understanding of the hot matter effect mechanisms affecting $\Upsilon$ states would require a better understanding of feed-down contributions at low $\pt$ and on the rapidity-dependence of the $\raa$.
In fact, recent measurements of the feed-down fraction to $\Upsilon$(1S) with $\pt>6$ GeV$/c$ by LHCb~\cite{Aaij:2012se} challenge the frequently used fraction of 50\% (based on higher $\pt$ measurements by CDF~\cite{Affolder:1999wm}), suggesting values closer to (or even below) 30\%. Such low feed-down fractions would imply suppression of directly produced $\Upsilon$(1S) at the LHC. Furthermore, sequential suppression models predict a minimum $\Upsilon$ $\raa$ value at mid-rapidity, while current data from ALICE~\cite{Abelev:2014nua} and CMS~\cite{Khachatryan:2016xxp} suggest a rather flat trend. A careful measurement of the rapidity dependence of $\Upsilon$ $\raa$ is, therefore, of great importance for the future.

In section II, we discussed dynamical models of quarkonium production that make contact with QCD. The common feature
of all these models is that quarkonium states are formed inside the QGP. Furthermore, the basic ingredient of all
these models are the force between the heavy $Q$ and $\bar Q$ that is related to the real part of the potential at $T > 0$,
and the stochastic force of the medium acting on the quark or anti-quark, which can be related to the imaginary part
of the potential at $T > 0$ or to the heavy-quark diffusion constant. In its simplest realization, such models
amount to Langevin dynamics of correlated $Q\bar Q$ pairs \cite{Young:2008he,Young:2009tj,Young:2011ug,Petreczky:2016etz}.
The stochastic forces of the medium will eventually destroy the correlations between $Q$ and $\bar Q$, but for a QGP with
finite life time some of the $Q\bar Q$ pairs will remain correlated and will form quarkonium states again. Which states will
be formed from the correlated $Q\bar Q$ pairs depends on their distribution in the relative distance at the time of the freeze-out or bound state formation. Calculations show that this distribution is peaked for small relative distances between
the $Q$ and $\bar Q$ \cite{Young:2008he,Petreczky:2016etz}, so ground state quarkonia are more likely to be formed than excited states.
Therefore, we could talk about sequential quarkonium formation in the QGP rather than sequential suppression. 

The above discussion was mostly focused on $Q\bar Q$ pairs that were correlated at the time of QGP formation or
earlier. This is the diagonal (re)generation of quarkonia. The off-diagonal (re)generation, where the quarkonium 
state is formed from initially uncorrelated $Q$ and $\bar Q$ can also be studied in Langevin dynamics \cite{Young:2009tj}.
This mechanism depends on the total number of heavy quark anti-quark pairs.
The Langevin dynamics of correlated $Q\bar Q$ pairs was embedded in the realistic hydrodynamic background and the
$J/\psi$ $\raa$ was calculated at RHIC \cite{Young:2008he,Young:2009tj} and LHC \cite{Young:2011ug}. The model
was able to explain the experimental results both at RHIC and LHC quite well. It was found that at RHIC the off-diagonal
recombination is small \cite{Young:2009tj}, while it is very important at LHC \cite{Young:2011ug}. This approach
can be considered as a more microscopic realization of statistical recombination of ref.~\cite{Zhao:2011cv}, in fact
it is a microscopic calculation of the correlation volume that enters in recombination models.
It should be noted that current lattice QCD calculations, e.g.\ Ref.~\cite{Kaczmarek:2016xix}, indicate a rapid charm quark equilibration on time scales similar to light quark equilibration times (${\approx}1$ fm/$c$) as well as charmonium dissociation temperatures close to $T_c$, and hence support off-diagonal regeneration.

The Langevin dynamics of correlated $Q\bar Q$ pairs is a valid approach in the limit of loosely bound $Q$ and $\bar Q$. So, it is clearly not applicable to ground state bottomonium. In that case, a full quantum treatment is required \cite{Katz:2015qja}. A possible way out is to treat the tightly bound $\Upsilon$(1S) as a distinct particle whose number is described by a rate equation, while other bottomonium states are treated using Langevin dynamics \cite{Petreczky:2016etz}. An effort to construct viable models that are based on the idea of bottomonium formation in the QGP is underway \cite{Petreczky:2016etz,Gossiaux:2016htk}.

%
The above debate is strongly linked to the interpretation of quarkonium suppression as a dissociation process. In the original sequential dissociation proposal~\cite{Matsui:1986dk}, it is assumed that bound states (color singlet) cannot be fully formed before the QGP is created.
It is important to stress that shrinking of the plasma equilibration time scales supports in fact the hypothesis of~\cite{Matsui:1986dk}. 
Even if $t_{\rm F}$ was found to be much smaller then $\tau_0$, some participants doubted that the correct physical picture should be the one of fully formed quarkonia entering a QGP, as the large fields existing in the pre-equilibrium phase would presumably prevent any binding of the $Q\bar{Q}$ pair, giving rise to correlated $Q\bar{Q}$ state at the time of thermalization, what could be appear as an effective increase of the vacuum formation time, which is also observed when an equilibrated QGP has been reached~\cite{Song:2013lov}.


\subsubsection{Role played by comovers}
Discussion on the role of comovers~\cite{Ferreiro:2014bia,Ferreiro:2012rq,Capella:2007jv} in the interpretation of $\raa$ results has taken place. One important point was that comovers are a rather abstract concept that scales with the particle multiplicity but without a clear connection to partons or hadrons.
While comovers seem to play a role to explain the suppression of excited states in pA collisions, their role in AA collisions (and also in pp) has extensively been debated, 
without reaching a firm conclusion. 
Since comovers are proportional to the particle multiplicity, it would be important to present nuclear modification results as a function of d$N_{\mathrm{ch}}$/d$\eta$ and seek for some universality features. In the comover model, the dissociation cross sections is chosen as free parameter. Suggestions were made to calibrate these on the width of the spectral functions deduced from lattice QCD.  
The comover models could help to constrain the hadronic break-up cross section of different quarkonium states. This
would have important implications for the modelling of quarkonium suppression by the hadronic medium at late
stages of heavy-ion collisions.


\section{Acknowledgments}
\noindent We are very thankful for the generous support by the Netherlands Organisation for Scientific Research (NWO), the National Institute for Subatomic Physics (Nikhef) in Amsterdam and the Lorentz fonds.

\noindent This work has been supported in part by the U.S Department of Energy through grant contract No.
DE-SC0012704 (P. Petreczky) and
DE-FG02-05ER41367 (S. A. Bass).

\noindent G. Aarts and C. Allton are supported by STFC (ST/ L000369/1). G. Aarts is supported by the Royal Society and the Wolfson Foundation.

\noindent N. Brambilla acknowledges the support by the Bundes\-ministerium f\"ur Bildung und Forschung (BMBF) under grant Verbundprojekt 05P2015 - ALICE at High Rate (BMBF-FSP 202) GEM-TPC Upgrade and Field theory based \mbox{investigations} of ALICE physics under grant No. 05P15WOCA1.

\noindent
This work has been supported in part by the DAAD Project 57210718 (E. Bratkovskaya and T. Song).

\noindent T. Dahms is supported by the Deutsche Forschungsgemeinschaft (DFG) Cluster of Excellence ``Origin and Structure of the Universe''.

\noindent S. K. Das acknowledges the support by the European Research Council under the grant no. 259684.

\noindent M. Djordjevic acknowledges support from Marie Curie IRG within the 7th EC Framework Programme (PIRG08-GA-2010-276913) and by the Ministry of Science of the Republic of Serbia under project numbers ON173052 and ON171004.

\noindent W. A. Horowitz appreciates the support by the South African National Research Foundation and the SA-CERN Collaboration.

\noindent M. G. Munhoz, A. C. Oliveira da Silva, A. A. P. Suaide and H. J. C. Zanoli acknowledge the support by Conselho Nacional de Desenvolvimento Cient'fico e Tecnol\'ogico (CNPq) and Funda\c{c}\~{a}o de Amparo ˆ Pesquisa do Estado de S\~{a}o Paulo (FAPESP).

\noindent The work of A. Mischke is supported by NWO, project number: 680-47-232, and the Dutch Foundation for Fundamental Research (FOM), project numbers: 12PR3083.

\noindent A. Rothkopf acknowledges support by the by the DFG Collaborative Research Centre "SFB 1225 (ISOQUANT)". 

\noindent L. Tolos acknowledges support from the Ramon y Cajal research programme, FPA2013-43425-P Grant from Ministerio de Economia y Competitividad (MINECO) and Spanish Excellence Network on Hadronic Physics FIS2014-57026-REDT from MINECO.


\bibliographystyle{epj}
\bibliography{lorentz_document}

\end{document}